\documentstyle[12pt,colordvi,epsfig]{article}
\newcommand{\be}{\begin{eqnarray}}
\newcommand{\ee}{\end{eqnarray}}

\newcommand{\nd}{\noindent}
\begin{document}
\begin{center}
\medskip
{\Large{\bf Quenching of Hadron Spectra in a chemically equilibrating
Quark-Gluon Plasma}}
\vskip 0.2in
\large{ B. K. Patra$^1$ and Madhukar Mishra$^2$}
\vskip 0.2in
\normalsize{$^1$ Dept. of Physics, Indian Institute of Technology Roorkee,
Roorkee 247 667, India\\
$^2$ Dept. of Physics, Banaras Hindu University,
Varanasi 221 005, India}
\end{center}

\vskip 1.2in

\begin{center}
{\bf Abstract}
\end{center}
Using the Fokker-Planck equation we have studied the drag co-efficient
$A(t)$ and the consequent shift $\Delta p_\perp (L)$ in
the transverse momentum due to collisional energy loss
of energetic partons while passing through a chemically equilibrating
quark-gluon plasma. Based on these we estimate the quenching factor
$Q(p_\perp)$ when the medium is undergoing longitudinal expansion
governed by master rate equations. In contrast to the case
of chemically equilibrated plasma investigated earlier by Mustafa and
Thoma~\cite{mus} we find less quenching
because our calculated $Q(p_\perp)$ is always greater
at all momenta. This result is attributed to the weak drag coefficient
operating during initial state interactions.

\vskip 0.9in

\nd PACS: 12.38.Bx; 24.85.+p

\newpage
\section{Introduction}
The inclusive yield of hadrons produced with high transverse momentum $p_\perp$
in Au+Au collisions at the Relativistic Heavy Ion Collider (RHIC)
has recently been shown to be significantly suppressed in comparison
with the cumulative yield of binary NN interactions. This effect, called
``jet quenching'', is considered an important signal for the production of
a quark-gluon plasma (QGP) and a vast literature ~\cite{review}-\cite{Baier}
already exists dealing with the phenomenological/theoretical aspects of the
said effect; see Appendix for a brief review.

Jet quenching is believed to occur due to the fact that hard partons
produced at the initial stage of the heavy ion collision lose energy as they
propagate through the fireball making $Q(p_\perp) <1$. Here the quenching factor
is experimentally defined  by the ratio
\begin{equation}
Q(p_\perp)= \frac{\Sigma^{{\rm med}}(p_\perp)}{\Sigma^{{\rm vac}}(p_\perp)};
\, \, \, \Sigma (p_\perp)= \frac{d^2 \sigma(p_\perp)}{d^2 p_\perp}
\end{equation}
where $\Sigma(p_\perp)$ is the inclusive hadron spectrum
at transverse momentum $p_\perp$, and the superscripts refer to the
medium and vacuum, respectively. The theoretical calculation of $Q$
invokes two important, mutually competing, mechanisms of the energy loss
described below.

The radiative mechanism~\cite{Baier} is caused by deceleration of the colour
charge accompanied by the bremsstrahlung of soft gluons; perturbative QCD allows
the determination of the soft gluon distribution $I(\omega)$ in terms of a
characteristic frequency $\omega_c$.

On the other hand, the collisional mechanism~\cite{BP}-\cite{Thoma4}
arises from the
elastic encounters with the other partons of the medium; here a QCD
motivated drag force permits the evaluation of the average momentum loss
$\Delta p_\perp$ suffered by the test parton over a specified  distance $L$. The
collisional loss theory has been recently applied by Mustafa and
Thoma~\cite{mus} (referred to as MT hereafter) assuming a temperature cooling
law relevant to chemically equilibrated QGP. The aim of the present paper
is to extend their theme to the case where the evolving fugacities
have not yet achieved chemical equilibrium.

For the sake of convenience Sec.2 below defines our notations and
briefly recapitulates the main formulae derived by MT. The details of our
numerical work are presented in Sec.3. Finally, physical interpretations
of the results along with some concluding remarks appear in Sec.4.

\section{Recapitulation of MT Theory~\cite{mus}}

{\em Step i)}  In the fireball rest frame a Taylor expansion of the hadron
spectrum $\Sigma$ is made~\cite{Baier} to rewrite (1) in the standard form
\be
\Sigma^{\rm{med}} (p_\perp) \approx \Sigma^{\rm{vac}} (p_\perp+\Delta E);
~~\Delta E=\int d \epsilon~\epsilon~D(\epsilon)
\ee
where $\epsilon$ is the random collisional energy loss over time span $t$, 
$E=E_0+\epsilon$ the random surviving energy, $D(\epsilon)$ the probability  
distribution that a parton loses the energy $\epsilon$, and $\Delta E$ the 
average energy loss in traversing the distance $L=c~t$.

{\em Step ii)} Let the instantaneous momentum of the leading parton moving in
the transverse $x$ direction be called $\vec{p}=\mid \vec{p} \mid \hat{e}_x$.
Assuming spatially uniform plasma a Boltzmann transport equation is set-up
for the distribution function $D(t,\vec{p})$, Landau's approximation
is made in the collision integral, and the resulting Fokker-Planck equation
is cast in the form
\begin{equation}
\frac{\partial D}{\partial t} = \frac{\partial}{\partial{\vec{p}}}
\left [ {\vec{{\cal T}}}_1(t,\vec{p}) D \right ] +
\frac{\partial^2}{\partial p^2}
\left [ {\cal T}_2(t,\vec{p}) D \right ].
\end{equation}
The transport coefficients $\vec{{\cal T}}_1$ and
${\cal T}_2$ are defined in terms of the net instantaneous  collision
rate $w (t,\vec{p},\vec{k})$, involving soft momentum transfer
$\vec{k}$, by
\begin{eqnarray}
\vec{{\cal T}}_1(t,p) &=& \int d^3k ~w(t, \vec{p}, \vec{k})\, \vec{k} \\
{\cal T}_2(t,\vec{p}) &=&\frac{1}{2} \int d^3k~w(t, \vec{p}, \vec{k})\, k^2 \\
w(t, \vec{p}, \vec{k}) &=& \sum_{j=q, {\bar q}, g} \gamma_j \int \
\frac{d^3q}{(2\pi)^3} f_j(t, \vec{q})~v_{\rm{rel}}~\sigma_j \, ,
\end{eqnarray}
where $j$ labels different species (quarks, antiquarks, gluons)
present in the plasma, $\gamma_j$ is the degeneracy factor, $f_j$ the
species' distribution function depending implicitly on the time
through the fugacity $\lambda_j(t)$ and temperature $T(t)$,
$v_{\rm{rel}}$  the relative speed between the test parton and
$j$-th species, and $\sigma_j$ the associated scattering cross
section.\\

{\em Step iii)}  Correspondence with classical dissipative motion tells that the
function $\vec{{\cal T}}_1$ scales like $\vec{p}$ and represents the rate of
energy loss $-dE/d{\vec{x}}$. Hence, a mean drag coefficient
$A(t)$ is constructed in one-dimensional notation via
\begin{equation}
{\cal A}(t) = \langle -\frac{1}{p}\, \, \frac{dE}{dL} \rangle
= \int d^3 p
\frac{-\frac{1}{p}\frac{dE}{dL}}{\exp{{\sqrt{p^2+m_g^2}/T }-1}}/
\int d^3 p \frac{1}{\exp{{\sqrt{p^2+m_g^2}/T} -1}}
\end{equation}

Since the QGP expected at RHIC and LHC is likely to be out of
chemical equilibrium it is necessary to investigate the energy loss in
this case \cite{Mustafa}.
Indeed, even away from chemical equilibrium,  dynamical screening
 remains operational within the HTL-resummed perturbation
theory. More explicitly, the collisional energy loss for a heavy quark
(mass $M$) propagating through a QGP parametrized in terms of the
distribution functions $\lambda_q n_F$  and $\lambda_g n_B$,
respectively,  where
$\lambda_{q,g}$ are the fugacity factors describing chemical
non-equilibrium, becomes~\cite{review}
\begin{equation}
  - \frac{dE}{dL}  = 2\alpha_s {\tilde m}^2_g~ \frac{(1+9/4)}{2}
\ln \left[ 0.920 \frac{\sqrt{ET}}{ \tilde m_g} \,
 2^{\lambda_q N_f/(12 \lambda_g + 2 \lambda_q N_f)}
                                \right].
\end{equation}
This expression \cite{review,Dirks}is valid for energetic quarks with 
$E \gg M^2/T$ and contains for $\lambda_q = \lambda_g = 1$ the original 
result of \cite{mus}. The screening mass parameter is
\begin{equation}   
\tilde m^2_g = 4 \pi \alpha_s (\lambda_g + \lambda_q N_f / 6 ) T^2 / 3 .
\end{equation}
The expression reduces to the expression given in Mustafa and Thoma paper
by putting $\lambda_g=\lambda_q=1$. It should be noted that the factor
$\frac{(1+9/4)}{2}$ arises due to averaging over the quark and gluon 
contributions.

Next, the analogy with Einstein's random walk relation is exploited to
identify the momentum-averaged diffusion coefficient
\begin{equation}
{\cal D}_F(t) = \langle {\cal T}_2(t,p) \rangle ={\cal A} T^2
\end{equation}
so that the evolution equation for the momentum-distribution  of the
Brownian particle becomes
\begin{equation}
\frac{\partial D}{\partial t} = {\cal A}~\frac{\partial }{\partial p} (p D)
+ {\cal D}_F ~\frac{\partial^2 D}{\partial p^2}
\end{equation}

{\em Step iv)}  Employing Fourier transform and method of characteristics the
above partial differential equation is solved analytically subject to
the initial condition $D(p,0)= \delta (p-p_0)$. The result in terms of
length and energy variable is
\begin{eqnarray}
D(L,~E) = \frac{1}{\sqrt{\pi {\cal W}(L)}}
\exp \left[ \frac{}{} - \frac{{\left( \frac{}{} E-E_0 \cdot B(L)
\frac{}{} \right)}^2}{{\cal W}(L)}
\right] \, ,
\end{eqnarray}
where
\begin{eqnarray}
B(L)=\exp \left(\frac{}{} -\int_{\tau_0}^L dt' ~{\cal A}(t') \frac{}{} \right) \, ,
\end{eqnarray}
and
\begin {equation}
{\cal W}(L) = 4\int_0^t dt'~ {\cal D}_F(t')~
\exp \left [ 2 \int_0^{t'}dt''~{\cal A}(t'')\, \right ]\,
B^2(L)\, .
\end{equation}

{\em Step v)}  The mean  energy $\langle E \rangle$ and average energy loss
$\Delta E$ are determined from
\begin{equation}
\langle \, E \, \rangle \, = \, \langle p \rangle \equiv \,
\int_0^\infty \, dE \, E \, D(L,E) \,
= \, E_0 \, B(L)
\end{equation}
and
\begin{eqnarray}
\Delta E\, = \Delta p \equiv E_0\, -\, \langle \, E \, \rangle
= \, E_0\left ( \frac{}{} \, 1\, - B(L) \frac{}{} \right ) \,
\end{eqnarray}
Eqs.(13, 14) can now be inserted into the basic expression (2) along
with the following high-energy parametrization for jet hadronization
at RHIC:
\begin{equation}
\Sigma^{{}^{\mbox{vac}}}(p_\perp)= \, {\mbox{const}} \left (1\, +\, \frac{p_\perp}{P_0}
\right )^{-\nu} \, \, \, ,
\end{equation}
with $\nu = 8$ and $P_0=1.75$ GeV.\\

{\em Step vi)}  Finally, the geometry of the head-on heavy ion collision is
accounted for by considering a cylindrical plasma of radius $R$ and the
test particle moving in the central rapidity region. If the latter was
created at the location $(r,\phi)$ in the transverse plane $z=0$ then it
travels a distance
\begin{equation}
L(r,\phi)= (R^2\, - \, r^2 \, \sin^2\phi \,)^{1/2}\, - \, r \, \cos \phi \, \,
\,
\end{equation}
before leaving the cylinder. Upon averaging (2) over the creation
configuration the effective quenching factor becomes
\begin{equation}
Q(p_\perp) = \frac{1}{2 \pi^2 R^2}  \int_0^{2\pi} d\phi~\int_0^R ~d^2r~
\Sigma^{{}^{\mbox{vac}}} (p_\perp+\Delta p) \frac{}{}/ \frac{}{}
\Sigma^{{}^{\mbox{vac}}} (p_\perp)
\end{equation}
which is amenable to direct computations.

\section{Numerical Application}

\nd {\em Mustafa-Thoma procedure}: MT assumed the QGP to expand longitudinally
according to Bjorken's boost-invariant hydrodynamics~\cite{bjor} so that
the temperature $T(t)$ on the transverse plane decreases according
to the scaling law
\begin{equation}
T(t)=T(t_0) {\left( \frac{t_0}{t} \right)}^{1/3},
\end{equation}
where $t_0$ the instant when the background plasma had just
attained local kinetic as well as chemical
equilibrium so that all partonic fugacities had become unity. For
an $A+B$ collision at RHIC, MT took
\be
t_0=0.3~{\rm{fm}}; ~~ T_0=0.5~{\rm{GeV}}
\ee
The momentum averaging in (7) was done using a Boltzmann distribution at
temperature $T(t)$ for gluons.\\

\nd {\em Our Procedure}: For the case of longitudinal expansion we know
that, in the early
stage of evolution, the plasma may achieve equilibrium thermally
but not yet chemically. Partonic reactions drive the system towards
chemical equilibration through the master rate equations~\cite{biro}
\be
&&\frac{\dot{\lambda_g}}{\lambda_g} + 3 \frac{\dot{T}}{T} +\frac{1}{\tau}
=R_3 \left(1-\lambda_g\right) - 2 R_2 \left(1-\frac{\lambda_g^2}{\lambda_q^2}
\right),\nonumber\\
&&\frac{\dot{\lambda_q}}{\lambda_q} + 3 \frac{\dot{T}}{T} +\frac{1}{\tau}
=R_2 \frac{a_1}{b_1}\left(\frac{\lambda_g}{\lambda_q} -
\frac{\lambda_q}{\lambda_g}\right),\nonumber\\
&&{\left(\lambda_g+\frac{b_2}{a_2} \lambda_q\right)}^{3/4} T^3 \tau =
{\rm{\mbox{const}}} \, ,
\ee
where the usual meaning of the symbols can be found in Ref.~\cite{biro}.
Their solutions subject to the initial conditions~\cite{hijing} (Table 1)
\begin{table}
\begin{center}
\caption{Different sets of initial conditions of the temperature,
  fugacities and parton number densities at $\tau_0=0.7$fm/c
  for RHIC and $\tau_0=0.5$fm/c for LHC.}
\begin{tabular}{lllllll}
\hline
\mbox{} & RHIC(1) & LHC(1) & RHIC(2) & LHC(2) & RHIC(3) & LHC(3) \\\hline
$T$(GeV)  & 0.55 & 0.82 & 0.55 & 0.82 & 0.4  & 0.72   \\
$\lambda_g$& 0.05 & 0.124& 0.2  & 0.496&0.53 & 0.761  \\
$\lambda_q$&0.008 & 0.02 & 0.032& 0.08 &0.083&0.118  \\
$n_g$(fm$^-3$)& 2.15 & 18 &8.6 &72 &8.6 &72  \\
$n_q$(fm$^-3$)&0.19 &1.573&0.76&6.29&0.76&6.29 \\
\end{tabular}
\end{center}
\end{table}
give the temperature and fugacities at general $t$, and these enter the
distributions $f_j(t,\vec{q})$ of various species written in (6). But the
momentum integration in (7) was performed employing an equilibrium
Bose-Einstein form at temperature $T(t)$.

\begin{figure}[h]
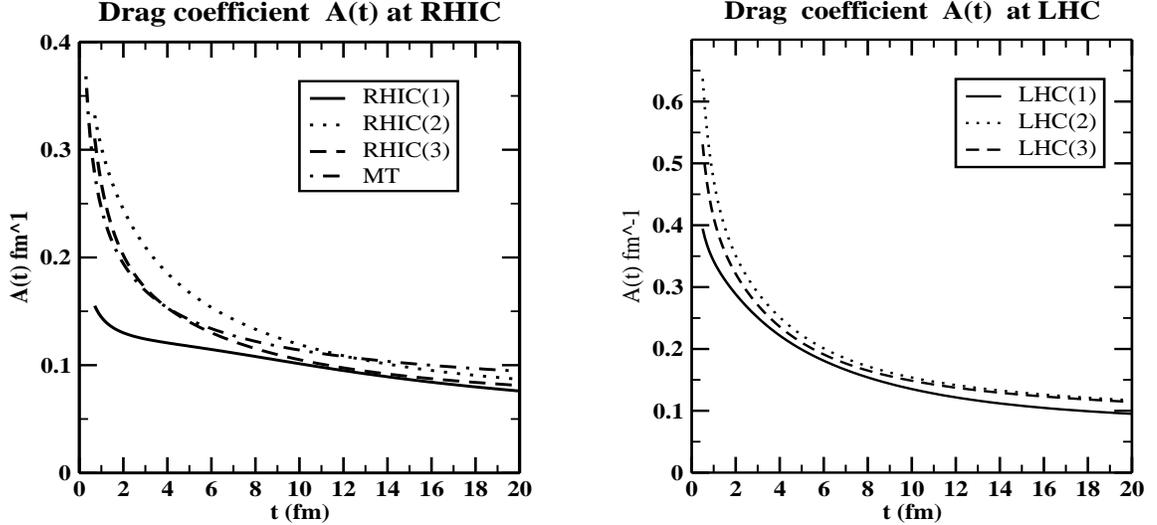

\mbox{
\epsfig{file=drag_rhic.eps,angle=0,height=7cm,width=7cm}
\hskip1cm
\epsfig{file=drag_lhc.eps,angle=0,height=7cm,width=7cm}}
\vskip0.5cm
\caption{The drag coefficient ${\cal A}(t)$ in an expanding QGP.
In the left panel, the solid, dotted and dashed line denotes chemically
equilibrating QGP corresponding to RHIC(1), RHIC(2) and RHIC(3) initial
conditions, respectively. The dashed-dotted line corresponds to MT's equilibrated
system. On the other hand right panel corresponds to LHC energy.
MT choose the initial values from eq.(21) with $\lambda_q= \lambda_g=1$ 
whereas latter takes the initial conditions from Table 1~\protect\cite{hijing}.}
\end{figure}

\nd {\em Computed Results}: Fig.1 plots the collisional drag coefficient
$A$ {\em versus} elapsed time $t$ for energetic gluons, in the QGP phase of the
expanding fireball, obtained from (7). The dashed-dotted line corresponds to
MT's equilibrated plasma described by the cooling law (20) whereas
other curves refer to our equilibrating plasma governed by the master
equations (22). Next, Fig.2 shows the relative energy loss 
$\Delta p_\perp/p_\perp$
versus the transverse distance $L$. Finally, Fig.3 displays the effective
quenching factor $Q$ versus the transverse momentum $p_\perp$. Logical
explanation of these  results is taken-up in the next section.

\begin{figure}[h]
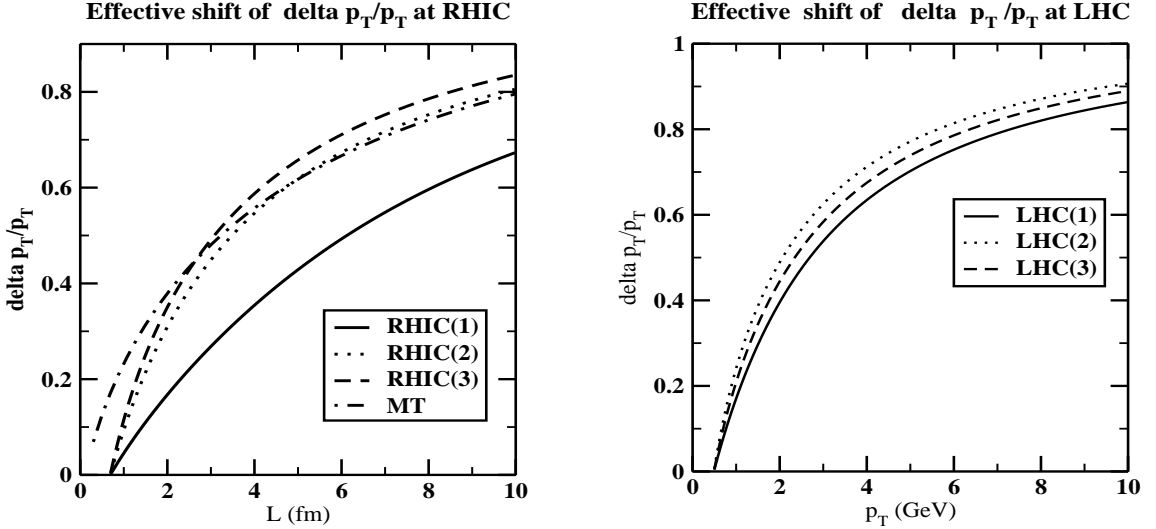

\vspace*{-0cm}
\mbox{
\epsfig{file=delpt_rhic.eps,angle=0,height=7cm,width=7cm}
\hskip1cm
\epsfig{file=delpt_lhc.eps,angle=0,height=7cm,width=7cm}}
\vskip 0.1in
\caption{The effective shift of the scaled
transverse momentum $\Delta p_\perp/p_\perp$ as a function
of traversed distance $L$. The notations of the curves are the same
as in Fig.1}
\end{figure}

\begin{figure}[h]
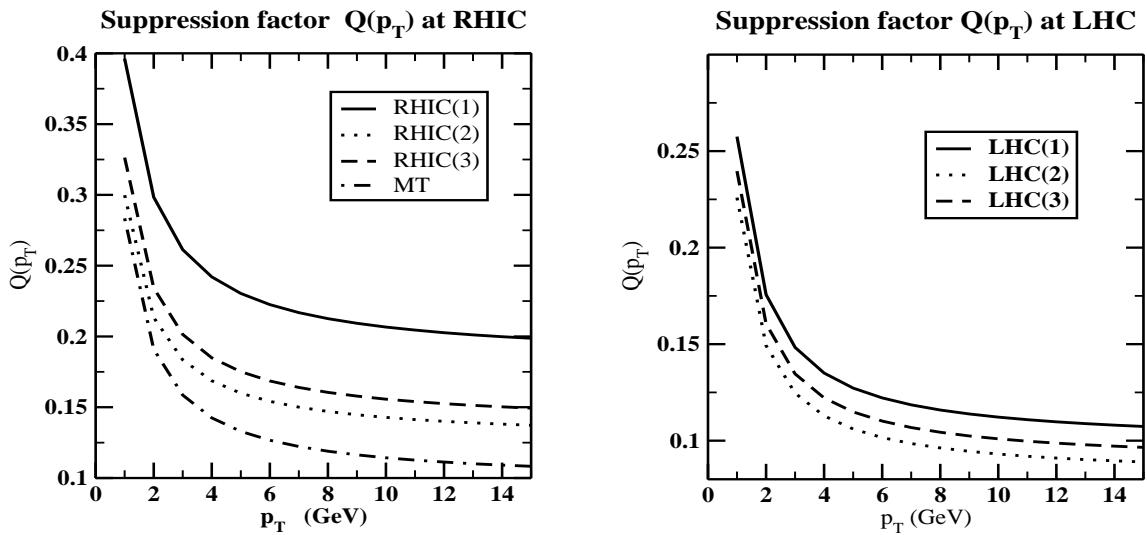

\vspace*{-3cm}
\mbox{
\epsfig{file=qpt_rhic.eps,angle=0,height=7cm,width=7cm}
\hskip1cm
\epsfig{file=qpt_lhc.eps,angle=0,height=7cm,width=7cm}}
\vskip 0.1in
\caption{The quenching factor $Q(p_\perp)$ as a function of
transverse momentum $p_\perp$. The notations of the curves are the same
as in Fig.1}
\end{figure}

\section{Interpretations and conclusions}
As is well known the distribution function $f_j(t,\vec{q})$ of the
$j$-th species in (6) grows with higher temperature $T(t)$ and fugacity
$\lambda_j(t)$. The numerical solution of (20) and (22), with
the initial conditions stated, reveals that Mustafa-Thoma fugacities
(and hence the transition rates $w(t,\vec{p},\vec{k})$) are about
an order of magnitude more than those in our work at all times of
experimental interest. Consequently, in the left panel of Fig.1 the drag 
coefficient $A^{\rm{MT}}$ is significantly greater than $A^{\rm{RHIC(1)}}$
; indeed the ratio $A^{\rm{MT}}/A^{\rm{RHIC(1)}} \approx 2$ at 
$t\approx 1~fm$. 
The reason for this difference lies in the fact that MT deals with a 
fully equilibrated system whereas RHIC(1) deals with the highly unequilibrated
system as is evidenced from their initial conditions. As a consequence, 
gluon density in MT system becomes larger than RHIC(1) resulting higher
value of the drag coefficient $A(t)$.
At large time (greater than 8 fm), our result 
does not differ much from MT result because as time elapses chemically
unequilibrated system (RHIC(1)) approaches towards chemical equilibrium 
making the two system alike. However, our 
calculated value is always less than $A^{\rm{MT}}$. Thus 
equilibrating QGP provides less drag force in comparison to equilibrated QGP. 

However for the illustration of the dependence of A(t) on initial
parton fugacities, we have calculated $A(t)$ at RHIC(2) and 
RHIC(3)~\cite{hijing} where the values of A(t) is not much different
from MT. This is due to the fact RHIC(2,3) is not much unequilibrated as
RHIC(1) as is evidenced by their initial gluon/quark fugacities or
gluon densities compared to RHIC(1). \\
In the right panel of Fig.1 the drag coefficient $A(t)$ at LHC energy 
also decreases with time. But our result at LHC energy is always somewhat 
greater than the value at RHIC. Thus the drag coefficient
increases with increase in the center-of-mass energy.

Next, in the left panel of Fig.2 the relative energy loss 
${\Delta p/p_\perp \mid}^{\rm{MT}}$
far exceeds ${\Delta p/p_\perp \mid}^{\rm{Our}}$ computed at sizable 
distances;
indeed their mutual ratio is about $1.4$ at $L\approx 7$ fm at RHIC(1) energy.
This happens because for higher integrand $A(t^\prime)$ the function
$B(L)$ becomes lower in (13, 16). Of course, in both cases $\Delta p$ scales
almost linearly with $p_\perp$ for a given $L$. In the right panel of the 
Fig.2 the relative energy loss ${\Delta p/p_\perp \mid}^{\rm{Our}}$ at LHC 
energy does not differ from the RHIC value. Thus increase in center-of-mass 
energy in going from RHIC to LHC the relative energy loss in equilibrating 
QGP will not be affected.

Finally, in the left panel of Fig.3, the quenching factor $Q(p_\perp)$ is 
stronger at lower $p_\perp$  in our calculation as well as MT calculation 
and gradually 
weakens at higher $p_\perp$ in agreement with experimental 
results~\cite{RHIC}. However, in our case, the quenching factor $Q(p_\perp)$ 
calculated for equilibrating QGP shows less suppression in comparison to 
quenching factor $Q(p_\perp)$ calculated for equilibrated QGP. This is again 
a reflection of the fact that, in contrast to MT equilibrated case, our 
energy loss $\Delta pi_\perp$ is much smaller over the whole assembly of 
lengths $L$ encountered in (18, 19). Moreover, the qualitative behavoiur 
is same in both the cases. In right panel of the Fig.3, the quenching factor 
$Q(p_\perp)$ shows stronger suppression at LHC energy in comparison to 
RHIC energy at lower values of the transverse momentum less than 
$p_\perp\approx 8~GeV$.

We end the paper with two concluding remarks. Although our assumptions
of chemically equilibrating QGP is more realistic than MT's assumption of
equilibrated plasma yet the equilibrating QGP will provide weak drag force for
the jets in comparison to equilibrated QGP. Thus most jet suppression occurs
after the equilibration of the QGP has been achieved. However, this 
{\it initial} state effect cannot be neglected. Therefore, jet quenching 
cannot be exactly called as a {\it final} state effect.

\section*{ACKNOWLEDGEMENTS}
BKP is thankful to V.J. Menon for the useful discussion during this work. 
BKP is also thankful to B. M\"uller, M. Thoma and M.G.Mustafa for their 
useful private communication.
BKP acknowledges the financial support to IIT Roorkee for the Faculty 
Initiation Grant and MM thanks to CSIR, New Delhi for his financial support.
\pagebreak

\pagebreak

\section*{Appendix: Jet quenching summarized}

The yield of hadron produced with high transverse momentum $p_\perp$ in
Au+Au collisions at RHIC has recently shown to be significantly suppressed
in comparison with the cumulative yield of NN collisions. This effect, so
called ``Jet quenching" was predicted to occur due to
energy loss suffered by hard scattered partons. The energy loss is expected
to occur due to interaction of the hard partons with the surrounding dense
medium.

There are two contributions to the energy loss of the patons in the
medium. One is due to the collisions among the partons in the medium and other
due to the radiation emitted by the decelerated colour charge i.e.,
bremsstrahlung of gluons.

In the initial stage of ultrarelativistic collisions energetic partons are
produced from hard collisions energetic partons are produced from hard
collisions between the partons of the nuclei. Receiving a large transverse
momentum, these partons will propagate through the fireball which might
consists of quark-gluon plasma phase for a transitional period of about
few fm/c. These high energy partons will manifest themselves as jets
leaving the fireball. These energy partons will loose energy due to
interaction of the hard partons with the fireball medium. Hence jet quenching
will result. The amount of jet quenching might depend on the state of the
fireball i.e., QGP or hot hadron gas, respectively. Therefore, jet quenching
has been proposed as the possible signature of the QGP formation.

It is difficult to measure the energy loss of the scattered partons directly
in heavy-ion collision because of the large multiplicity of the emitted
hadrons makes it almost impossible to isolate the resulting jet by the
kinematic cut. However, this energy loss of the hard partons may affect
the equivalent loss of the energy of the leading hadrons
produced it its fragmentation. this is what has been observed at RHIC
experiment~\cite{}.
Preliminary data from run 2 at RHIC Au+Au at $\sqrt{s_{NN}}=200$ GeV
confirm the effect observed in run 1 and its interpretation as jet quenching.
For pions with $p_\perp \approx$ 5 GeV/c the measured suppression factor is
about $1/5$.
\end{document}